\documentclass[sigconf]{acmart}
\AtBeginDocument{%
  }

\setcopyright{acmlicensed}
\copyrightyear{2025}
\acmYear{2025}
\setcopyright{acmlicensed}\acmConference[FSE Companion '25]{33rd ACM International Conference on the Foundations of Software Engineering}{June 23--28, 2025}{Trondheim, Norway}
\acmBooktitle{33rd ACM International Conference on the Foundations of Software Engineering (FSE Companion '25), June 23--28, 2025, Trondheim, Norway}
\acmDOI{10.1145/3696630.3730572}
\acmISBN{979-8-4007-1276-0/2025/06}

\usepackage{algorithmic}
\usepackage{graphicx}
\usepackage{textcomp}
\usepackage{xcolor}
\usepackage{enumitem} 

\usepackage{colortbl}
\usepackage{hyperref}

\usepackage{subfig}   
\usepackage{amsmath}
\usepackage{listings}
\usepackage{pifont}
\usepackage{bbding}
\usepackage{csquotes}

\usepackage[many]{tcolorbox}
\usepackage{booktabs}
\usepackage{cleveref}
\usepackage{url}
\usepackage{xspace}

\usepackage{multirow}
\usepackage{makecell}

\begin{document}

\title{Aligning Core Aspects: Improving Vulnerability Proof-of-Concepts via Cross-Source Insights}

\author{Lingxiao Wang}
\orcid{0009-0006-9364-8851}
\affiliation{%
  \institution{Tianjin University}
  \country{China}
}
\email{wlx\_0613@tju.edu.cn}

\author{Wenjing Dang}
\orcid{0009-0006-2169-8024}
\affiliation{%
  \institution{Tianjin University}
  \country{China}}
\email{dangwenjing@tju.edu.cn}

\author{Mengyao Zhao}
\orcid{0009-0001-3168-7787}
\affiliation{%
  \institution{Tianjin University}
  \country{China}}
\email{mengyaozhao@tju.edu.cn}

\author{Yue Wang}
\orcid{0009-0009-1217-8926}
\affiliation{%
  \institution{Tianjin University}
  \country{China}}
\email{16622536761@163.com}

\author{Xianzong Wu}
\orcid{0009-0003-8539-4757}
\affiliation{%
  \institution{Tianjin University}
  \country{China}}
\email{wxz\_buta@tju.edu.cn}

\author{Sen Chen}
\authornote{Sen Chen is the corresponding author.}
\orcid{0000-0001-9477-4100}
\affiliation{%
  \institution{Nankai University}
  \country{China}}
\email{senchen@nankai.edu.cn}

\renewcommand{\shortauthors}{Lingxiao et al.}

\begin{abstract}
For vulnerabilities, Proof-of-Concept (PoC) plays an irreplaceable role in demonstrating the exploitability. PoC reports may include critical information such as specific usage, test platforms, and more, providing essential insights for researchers. However, in reality, due to various PoC templates across PoC platforms, PoC reports extensively suffer from information deficiency, leading the suboptimal quality and limited usefulness. Fortunately, we found that information deficiency of PoC reports could be mitigated by the completion from multiple sources given the same referred vulnerability.

In this paper, we conduct the first study on the deficiency of information in PoC reports across public platforms. We began by collecting 173,170 PoC reports from 4 different platforms and defined 8 key aspects that PoCs should contain. By integrating rule-based matching and a fine-tuned BERT-NER model for extraction of key aspects, we discovered that all PoC reports available on public platforms have at least one missing key aspect. Subsequently, we developed a multi-source information fusion method to complete the missing aspect information in PoC reports by leveraging CVE entries and related PoC reports from different sources. 
Finally, we successfully completed 69,583 PoC reports (40.18\% of all reports).
\end{abstract}



\received{9 March 2025}
\received[accepted]{10 April 2025}

\maketitle

\section{Introduction}

In the field of software security, vulnerability reports are the primary means by which researchers learn about security flaws. 
Proof-of-Concept (PoC) plays an indispensable role by offering practical exploitation methods and code examples. PoC reports usually involve a description, affected software, and additional information that helps gain a deeper understanding of the vulnerabilities at hand. PoCs assist developers in pinpointing the vulnerabilities, thereby facilitating the remediation process and evaluation of exploitability~\cite{bozorgi2010beyond, wang2018revery, suciu2022expected}. 


Furthermore, the content of PoC reports, including usage details and testing platforms, has been analyzed in a few studies~\cite{kwon2021octopocs,mu2018understanding}, which emphasize the significant impact of such information on the applicability of PoCs. Mu et al.~\cite{mu2018understanding} specifically highlighted the crucial role that PoC reports play in the process of vulnerability reproduction.
However, despite their importance, the quality of PoC reports varies greatly, as noted in recent research~\cite{QualysSu67:online}. The type and completeness of information differ significantly across platforms, resulting in frequent omissions of critical details. This inconsistency reduces the overall usability of public PoC reports. Furthermore, there is still a lack of comprehensive research and effective methodologies to systematically address the issue of missing information in vulnerability PoCs.


To bridge the gap, we conducted the first empirical study to understand the information deficiency of PoC reports across public vulnerability report platforms. Specifically, this paper investigates the prevalence of missing information in PoC reports from various sources and proposes a multi-source information completion approach to effectively rectify these gaps.
First, we collate PoC reports from multiple public repositories and conduct a series of analytical studies. \ding{172} \textbf{Evaluating Information Deficiency of PoC Reports}. We identified 8 key aspects that PoC reports should encompass. For each aspect, we develop specific extraction approaches to accurately identify and extract information. 
We found on average only $56\%$ of aspects were present in PoC reports, indicating a significant absence of crucial data.
\ding{173} \textbf{Completion of PoC Aspects Based on Multiple Sources}. To remedy the missing information, we cross-reference PoCs with corresponding CVE entries, facilitating a linkage between different sources of PoC reports. We categorized the PoCs into code-based and text-based, designed methods to complete PoC reports with related CVE entries and other PoC reports based on aspects. Our experiments demonstrate the effectiveness of these completion strategies.

\section{Background}

\subsection{Diverse PoC Submission Templates for Multiple Platforms}
PoC reports are published at diverse sources, including platforms like ExploitDB~\cite{EDB} and CXSecurity~\cite{CXSECURI98:online}, where security researchers, from independents to professional firms, share PoCs for various products and vulnerabilities. While both platforms aim to facilitate security issue resolution, they differ in PoC handling, requiring distinct submission templates and mandatory information.
Incomplete templates or submissions that do not fully adhere to the specified templates can result in missing information. This issue is particularly prevalent on platforms with optional template requirements, such as Packet Storm~\cite{PacketSt14:online}. Thus, single-source PoC reports often suffer from severe information deficiencies, which not only diminish their usability but also hinder a comprehensive understanding of the relationship between the PoC and the associated vulnerability.


\subsection{Motivating Example}
To address the information gaps in PoC reports, we utilize complementary reports from different sources. For instance, a PoC from Packet Storm related to CVE-2003-0264 provides minimal details beyond execution code, lacking critical information such as affected software versions. In contrast, an associated PoC from ExploitDB for the same CVE~\cite{EDB638:online} includes comprehensive details like the author, operating platform, and execution results, enriching the dataset.


\section{Data Preparation}
\subsection{Data Collection}
To gather the most diverse and reliable open-source PoC data possible, we first sourced links tagged with ``Exploit'' from the ``References to Advisories, Solutions, and Tools'' section on the NVD. Links associated with the vulnerabilities are tagged according to their relevance, with links labeled as ``Exploit'' indicating relevance to vulnerability exploitation. We crawled all links with the ``Exploit'' tag and traced their origins, eventually selecting the top ten data sources based on reference frequency for preliminary inclusion, and finally select four data sources with higher-quality and more structured PoC content: ExploitDB~\cite{EDB}, Packet Storm~\cite{PacketSt14:online}, Seebug~\cite{Seeb31:online}, and CXSecurity~\cite{CXSECURI98:online}.
We employed techniques such as web scraping and the use of official data, tailoring data extraction methods specifically for each of the aforementioned 4 sources. This process involved preliminary data cleaning to remove erroneous entries that did not contain PoC information. 
Ultimately, over 200 man-hours were expended to collect a total of 173,170 PoC reports from these 4 data sources.

\subsection{Categorizing PoC into Code and Text}
\noindent{\textbf{Motivation for Categorizing.}} Given the diversity of PoC report formats, employing a uniform processing approach could lead to a substantial loss of format-specific information. We categorize PoCs into code-based, text-based, and other forms to preserve the uniqueness of each type. Code-based PoCs, such as the Python code PoC~\cite{SLMailRemote:online}, involve parsing based on syntactic and functional attributes. Text-based PoCs consist of narrative descriptions, while other forms include non-textual media like images or videos, which require different identification methods~\cite{suciu2022expected}.

\noindent{\textbf{Categorization Approach.}} Our analysis primarily focuses on code-based and text-based PoCs, as PoCs in non-textual forms are less prevalent. We distinguish these types by utilizing regular expressions to detect programming languages, such as C/C++, HTML, Java, Python, Perl, PHP, Python, Ruby, and Shell, commonly used in PoCs. If a PoC does not match these language patterns, it is classified as text-based~\cite{suciu2022expected}.

\section{Empirical Study}

\begin{figure}[]
  \centering
  \includegraphics[width=0.48\textwidth]{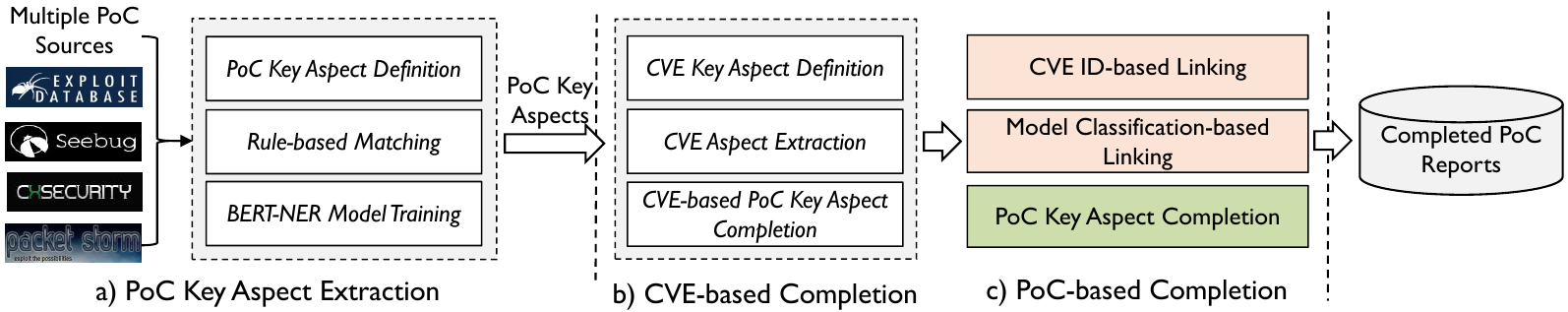}
  \caption{Overview of our study.}
  \label{fig:overall.pdf}
\end{figure}

The overview of our study is shown in~\Cref{fig:overall.pdf}.

\subsection{Evaluating Information Deficiency of PoC Reports}
\subsubsection{Key Aspects of PoC}
To standardize the information in PoC reports and ensure uniform processing, we identified eight key aspects crucial for understanding and utilizing them effectively. Drawing from research by Mu et al.~\cite{mu2018understanding}, which underscores the importance of certain elements in reproducing vulnerabilities, we categorized these aspects into two groups: Exploit and Basic. 
\textbf{Exploit Category} represents the exploit value of PoC reports. These 4 aspects are indispensable for the usability of a PoC, including \textit{Trigger Step}, \textit{Verification Oracle}, \textit{Test Platform}, \textit{Software Version}. \textbf{Basic Category} represents the reference value, including \textit{Title}, \textit{Author}, \textit{Publish Time} and \textit{Reference}.
These eight key aspects, termed the \textbf{Key Aspects} of a PoC report, furnish layered information critical for the accurate reproduction of vulnerabilities and essential references for further exploration.

\subsubsection{Extraction of Key Aspects}

Based on the 8 key aspects of PoC reports, we have developed extraction methods incorporating rule-based matching and NLP model-based approaches. 

 \textbf{\textit{(1) Rule-based Matching}}: We have established specific rules for \textit{Trigger Step}, \textit{Verification Oracle}, and \textit{Reference} based on observations. For the \textit{Trigger Step}, we have defined keywords such as \texttt{``steps''}, \texttt{``reproduce''}, and \texttt{``complie with''}, along with regular expressions to match potential indicators and lists of numerical and alphabetical steps that may appear in a PoC report. For the \textit{Verification Oracle}, we have set keywords like \texttt{``expected output''} and \texttt{``PoC output''} to identify possible results from program execution. \textit{Reference} are generally in the form of links, so we utilized a regular expression that matches the structure of web URLs to extract.

\textbf{\textit{(2) BERT-NER Model}}: Referring to the work of Guo et al~\cite{guo2022detecting}, for extracting less structured and more nuanced information such as \textit{Test Platform}, \textit{Software Version}, \textit{Title}, \textit{Author}, and \textit{Publish Time}, we utilized a BERT-NER model~\cite{devlin2018bert}. This model was trained on a subset of 2,500 manually annotated PoCs, using the BIO tagging scheme to identify these aspects effectively. The training involved 2,000 PoCs, with the remaining 500 serving as a test set. Fine-tuning parameters included a learning rate of 2e-5, ten epochs, weight decay of 0.01, label smoothing of 0.1, and Cross-Entropy Loss as the loss function. The training utilized an NVIDIA RTX 3090 GPU, ensuring efficient processing and robust model performance for accurate aspect extraction.

The overall precision and recall for extracting key aspects of PoC reports are 0.88 and 0.81 respectively. For PoC aspects, the precision and recall for \textit{Publish Time}, \textit{Reference}, \textit{Trigger Step}, and \textit{Verification Oracle} all exceeded 0.9, which ensures great extraction results for each aspect. 

\subsection{Completion of PoC Aspects Based on Multiple Sources}
The completion rules comprise two categories: \ding{172} Completion using CVE entries and \ding{173} Completion using PoC reports.
\subsubsection{PoC Completion Approach by Aspects of CVE Entries}

\begin{table}[]
\caption{Count of completed PoC reports and aspects by CVE entries.}
\label{tab:cvecompleteness}
\setlength{\tabcolsep}{5pt}
\resizebox{1\linewidth}{!}{\begin{tabular}{lcc|cc|cc}
\hline
   \multicolumn{3}{c|}{}          & \multicolumn{2}{c|}{\textbf{Test Platform}}                    & \multicolumn{2}{c}{\textbf{Software Version}}                  \\ \hline
\textbf{Sources} & \textbf{CVE}       & \textbf{PoC w/ CVE}  & \textbf{PoCs} & \textbf{Aspects} & \textbf{PoCs} & \textbf{Aspects} \\
\hline
\rowcolor[HTML]{D9E1F4}
Packet St.     & 7,465                 & 8,478                                           & 1,203                         & 9,605                            & 7,381                         & 19,455                           \\
ExploitDB              & 24,309                & 26,517                                             & 1,784                         & 8,574                            & 18,790                        & 39,728                           \\
\rowcolor[HTML]{D9E1F4}
Seebug           & 11,861                & 14,400                                           & 666                          & 2,215                            & 9,292                         & 20,158                           \\
CXSecurity       & 18,334                & 15,812                                           & 1,499                         & 6,387                            & 12,285                        & 26,119                           \\ \hline
\rowcolor[HTML]{D9E1F4}
\textbf{Overall}          & \textbf{34,024}                & \textbf{65,207}                                          & \textbf{5,152}                         & \textbf{26,781}                           & \textbf{47,748}                        & \textbf{105,460}                          \\ \hline
\end{tabular}
}
\begin{enumerate}
    \item  \textit{Poc w/ CVE} refers to PoC reports with CVE IDs. 
\end{enumerate}
\end{table}

To enhance the connectivity and completeness of PoC reports, we utilized CVE IDs as a unifying reference. 
We developed source-specific methods to extract CVE IDs from PoCs, recognizing that their locations vary across platforms—for example, ExploitDB houses them in a fixed field. This targeted extraction approach helps avoid the false positives associated with naive pattern matching. Our collection identified 34,024 unique CVE IDs linked to 65,207 PoC reports.
Research on CVE entries~\cite{sun2023aspect} shows they contain comprehensive details like affected software and versions, platforms, and vulnerability types. By mapping these CVE details to the corresponding PoC reports, we can enrich PoCs lacking specific information. This is critical as PoC reports often serve to validate and reproduce vulnerabilities. We specifically correlated \textit{Software Version} and \textit{Test Platform} from PoCs with the \textbf{Version} and \textbf{Platform} from CVE entries, using \textbf{Product} merely for verifying associations between CVE entries and PoC reports. 


We employed CVE Aspects to complete PoC Aspects. For the \textbf{Version}, our completion strategy adds version information from CVE entries related to the same CVE ID, based on the verification of the \textbf{Product} field. Since one CVE may affect multiple software, the \textbf{Version} entries are mapped in a 1-to-n fashion. A PoC, however, is typically written for one specific version. Therefore, if the original PoC lacks version information, we directly complete its \textit{Software Version} aspect. If it contains a partial version list, we append the missing versions for reference. 
The \textbf{Version} field in CVE Aspects corresponds to the \textit{Software Version} in PoC Aspects, while the \textbf{Platform} field maps to the \textit{Test Platform} in PoC reports. Completeness is evaluated by comparing these two aspects in each PoC against the corresponding CVE data. This includes both missing and incomplete cases in the original PoC.


The results of our completion approach are summarized in~\Cref{tab:cvecompleteness}. In total, \textbf{47,748} PoC reports were completed, covering \textbf{132,241} aspects. Notably, completions for \textit{Software Version} far exceed those for \textit{Test Platform}. Specifically, \textbf{6,482} PoCs and \textbf{34,783} aspects were completed for \textit{Test Platform}, while \textbf{58,222} PoCs and \textbf{133,493} aspects were completed for \textit{Software Version}.
This gap arises because not all CVE entries include \textbf{Platform} data, whereas all include affected \textbf{Product} and \textbf{Version} fields. Furthermore, a single CVE often involves multiple software and versions, making it rare for a single PoC to include all relevant version information—resulting in more opportunities for completion.
In conclusion, the experiment confirms the practicality and effectiveness of leveraging CVE entries to enrich PoC reports.

\subsubsection{PoC Completion Approach by other Related PoCs}

To establish connections among PoCs from different data sources, we employed two distinct methods. 

\noindent{\textbf{Based on CVE ID.}} 
For PoCs linked by a common CVE ID, we leverage the CVE as a marker to associate PoCs from varied sources, as these often pertain to the same vulnerability and share details like triggering mechanisms and exploitation logic. To enhance the completeness of PoC reports, we identify gaps in information that could be mutually filled based on their similarities.
Nevertheless, PoC reports vary significantly due to different programming languages, impacting the applicability of shared information across reports. We address this by clustering PoC reports based on both CVE ID and the programming languages used, acknowledging that even similar vulnerabilities might be exploited differently across various reports. To ensure precision in information completion, we focus on PoCs with high similarity.
\begin{itemize}[leftmargin=7pt]
\item \textbf{Code-Based}. We utilize a token-based method to measure code similarity across languages, including C/C++, Ruby, Python, PHP, JavaScript, Java, Perl, HTML, and Shell. This involves tokenizing the code and calculating token frequencies to establish a frequency dictionary for each PoC, with similarity assessed via Cosine Similarity~\cite{ye2011cosine}.
\item \textbf{Text-Based}. 
We employ the Word2Vec~\cite{mikolov2013efficient} to generate text vectors, comparing them using Cosine Similarity to gauge the similarity in described exploit processes. This methodological approach enables a structured and quantitative comparison of PoCs, facilitating the targeted completion of information within PoCs.
\end{itemize}

\noindent{\textbf{Based on Model classification.}} 
Linking PoCs without CVE IDs poses a challenge; hence, we developed a methodology to associate PoCs affecting the same vulnerability based on their software, version, and similarity in content. Initial analysis revealed that PoCs for the same vulnerability often impact the same software, albeit potentially different versions, and share similar titles and content.

To establish whether two PoCs target the same software, we compared the extracted software names for exact matches. Subsequently, we used the similarity of titles and PoC content as indicators of identical vulnerabilities. For this purpose, we trained a BERT classification model~\cite{devlin2018bert} to differentiate whether two PoCs are associated with the same vulnerability. The model inputs include the titles and content of the PoCs. We constructed the training dataset using pairs of PoCs: those with the same CVE ID served as positive samples, and those with different CVE IDs as negative samples. Given that PoCs with CVE IDs comprise only 13.8\% of our dataset, we selected 600 positive and 5,400 negative samples, maintaining an 8:1:1 training split. The model achieved an accuracy of 97.3\% on the test set, validating the effectiveness of our approach in linking PoCs based on content similarity and software identity.

\subsubsection{Results of Completion by Related PoCs}
\begin{figure}[]
  \centering
  \includegraphics[width=0.44\textwidth]{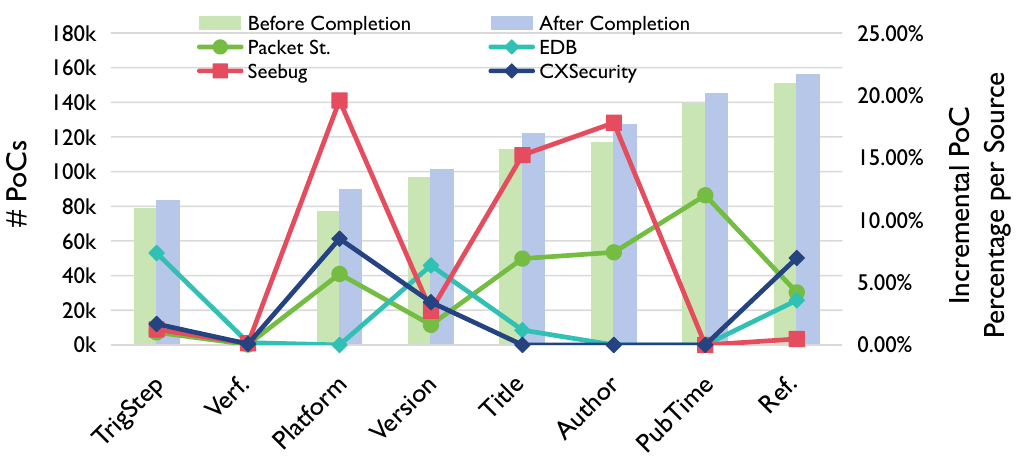}
  \caption{Overall completion results.}
  \label{fig:finresult.pdf}
\end{figure}

To ensure that paired PoCs accurately reflect the same exploit process, we conducted threshold-setting experiments and established distinct thresholds for PoCs linked by CVE ID. For code-based PoCs, the threshold was set at 0.5, while for text-based PoCs, we set it at 0.95. Utilizing these specific thresholds and completion rules, we carried out an information completion exercise across multiple PoC sources. The outcomes of this exercise are illustrated in~\Cref{fig:finresult.pdf}. The displayed bar graph details the number of PoCs containing each aspect both before and after the completion process, and the line graph highlights the incremental completions sourced from various databases.

We completed a total of 27,901 PoC reports across four datasets, focusing on those originally lacking specific aspect information. Furthermore, we completed 4,613 to 12,640 PoC reports by completing seven different aspects that were originally missing from these PoCs. Due to the originally small number of \textit{Verification Oracles}, the number of PoCs that could be completed was limited to just 184. For the significantly lacking \textit{Trigger Step}, we completed over 5,002 PoCs. Additionally, for the crucial aspects of \textit{Test Platform} and \textit{Software Version}, which are vital for reproducing vulnerabilities, we completed more than 12,640 and 4,613 PoC reports, respectively.



From the completion results across different sources, it is clear that the four datasets showed varying degrees of improvement across the eight aspects. Notably, Seebug exhibited significant enhancement, with completions for \textit{Test Platform} increasing by over 20\%, and both \textit{Title} and \textit{Author} exceeding 15\%. These gains are largely due to substantial missing information in the original PoCs.
In summary, our approach of leveraging related PoCs for aspect completion has proven highly effective, significantly enriching the quality and completeness of PoC information across multiple sources.

\section{Conclusion}
In this study, we collected PoC reports from 4 different sources and defined 8 key aspects of PoCs, revealing that on average only 56\% of these aspects are present across sources. We used CVE entries linked via CVE IDs to supplement missing aspects in PoCs, successfully completing 58,222 PoCs with 133,493 additional aspects.
We utilized PoCs from various sources for information completion, employing methods based on CVE IDs and model recognition to establish connections between different PoCs. To ensure the reliability of our approach, we further classified PoCs into text and code types, assigning specific similarity calculation methods to each type. Ultimately, this strategy enabled the successful completion of 27,901 PoC reports using related PoCs from different sources. In total, we completed 69,583 PoC reports, demonstrating the feasibility of this method and significantly enhancing the reliability of PoC data.

\bibliographystyle{ACM-Reference-Format}
\bibliography{ref}


\begin{thebibliography}{17}


\ifx \showCODEN    \undefined \def \showCODEN     #1{\unskip}     \fi
\ifx \showISBNx    \undefined \def \showISBNx     #1{\unskip}     \fi
\ifx \showISBNxiii \undefined \def \showISBNxiii  #1{\unskip}     \fi
\ifx \showISSN     \undefined \def \showISSN      #1{\unskip}     \fi
\ifx \showLCCN     \undefined \def \showLCCN      #1{\unskip}     \fi
\ifx \shownote     \undefined \def \shownote      #1{#1}          \fi
\ifx \showarticletitle \undefined \def \showarticletitle #1{#1}   \fi
\ifx \showURL      \undefined \def \showURL       {\relax}        \fi
\providecommand\bibfield[2]{#2}
\providecommand\bibinfo[2]{#2}
\providecommand\natexlab[1]{#1}
\providecommand\showeprint[2][]{arXiv:#2}

\bibitem[Qua(2024)]%
        {QualysSu67:online}
 \bibinfo{year}{2024}\natexlab{}.
\newblock \bibinfo{title}{{Qualys Survey of Top 10 Exploited Vulnerabilities in 2023 | Qualys Security Blog}}.
\newblock \bibinfo{howpublished}{\url{https://blog.qualys.com/qualys-insights/2023/09/26/qualys-survey-of-top-10-exploited-vulnerabilities-in-2023}}.
\newblock


\bibitem[Bozorgi et~al\mbox{.}(2010)]%
        {bozorgi2010beyond}
\bibfield{author}{\bibinfo{person}{Mehran Bozorgi}, \bibinfo{person}{Lawrence~K Saul}, \bibinfo{person}{Stefan Savage}, {and} \bibinfo{person}{Geoffrey~M Voelker}.} \bibinfo{year}{2010}\natexlab{}.
\newblock \showarticletitle{{Beyond heuristics: learning to classify vulnerabilities and predict exploits}}. In \bibinfo{booktitle}{\emph{Proceedings of the 16th ACM SIGKDD international conference on Knowledge discovery and data mining}}. \bibinfo{pages}{105--114}.
\newblock


\bibitem[Cxsecurity(2024)]%
        {CXSECURI98:online}
\bibfield{author}{\bibinfo{person}{Cxsecurity}.} \bibinfo{year}{2024}\natexlab{}.
\newblock \bibinfo{title}{{CXSECURITY.COM Free Security List}}.
\newblock \bibinfo{howpublished}{\url{https://cxsecurity.com/}}.
\newblock


\bibitem[Database(2024)]%
        {EDB638:online}
\bibfield{author}{\bibinfo{person}{Exploit Database}.} \bibinfo{year}{2024}\natexlab{}.
\newblock \bibinfo{title}{{Seattle Lab Mail (SLmail) 5.5 - POP3 'PASS' Remote Buffer Overflow (1)}}.
\newblock \bibinfo{howpublished}{\url{https://www.exploit-db.com/exploits/638}}.
\newblock


\bibitem[Devlin et~al\mbox{.}(2018)]%
        {devlin2018bert}
\bibfield{author}{\bibinfo{person}{Jacob Devlin}, \bibinfo{person}{Ming-Wei Chang}, \bibinfo{person}{Kenton Lee}, {and} \bibinfo{person}{Kristina Toutanova}.} \bibinfo{year}{2018}\natexlab{}.
\newblock \showarticletitle{{Bert: Pre-training of deep bidirectional transformers for language understanding}}.
\newblock \bibinfo{journal}{\emph{arXiv preprint arXiv:1810.04805}} (\bibinfo{year}{2018}).
\newblock


\bibitem[Guo et~al\mbox{.}(2022)]%
        {guo2022detecting}
\bibfield{author}{\bibinfo{person}{Hao Guo}, \bibinfo{person}{Sen Chen}, \bibinfo{person}{Zhenchang Xing}, \bibinfo{person}{Xiaohong Li}, \bibinfo{person}{Yude Bai}, {and} \bibinfo{person}{Jiamou Sun}.} \bibinfo{year}{2022}\natexlab{}.
\newblock \showarticletitle{Detecting and augmenting missing key aspects in vulnerability descriptions}.
\newblock \bibinfo{journal}{\emph{ACM Transactions on Software Engineering and Methodology (TOSEM)}} \bibinfo{volume}{31}, \bibinfo{number}{3} (\bibinfo{year}{2022}), \bibinfo{pages}{1--27}.
\newblock


\bibitem[Kwon et~al\mbox{.}(2021)]%
        {kwon2021octopocs}
\bibfield{author}{\bibinfo{person}{Seongkyeong Kwon}, \bibinfo{person}{Seunghoon Woo}, \bibinfo{person}{Gangmo Seong}, {and} \bibinfo{person}{Heejo Lee}.} \bibinfo{year}{2021}\natexlab{}.
\newblock \showarticletitle{{OCTOPOCS: automatic verification of propagated vulnerable code using reformed proofs of concept}}. In \bibinfo{booktitle}{\emph{2021 51st Annual IEEE/IFIP International Conference on Dependable Systems and Networks (DSN)}}. IEEE, \bibinfo{pages}{174--185}.
\newblock


\bibitem[Mikolov et~al\mbox{.}(2013)]%
        {mikolov2013efficient}
\bibfield{author}{\bibinfo{person}{Tomas Mikolov}, \bibinfo{person}{Kai Chen}, \bibinfo{person}{Greg Corrado}, {and} \bibinfo{person}{Jeffrey Dean}.} \bibinfo{year}{2013}\natexlab{}.
\newblock \showarticletitle{{Efficient estimation of word representations in vector space}}.
\newblock \bibinfo{journal}{\emph{arXiv preprint arXiv:1301.3781}} (\bibinfo{year}{2013}).
\newblock


\bibitem[Mu et~al\mbox{.}(2018)]%
        {mu2018understanding}
\bibfield{author}{\bibinfo{person}{Dongliang Mu}, \bibinfo{person}{Alejandro Cuevas}, \bibinfo{person}{Limin Yang}, \bibinfo{person}{Hang Hu}, \bibinfo{person}{Xinyu Xing}, \bibinfo{person}{Bing Mao}, {and} \bibinfo{person}{Gang Wang}.} \bibinfo{year}{2018}\natexlab{}.
\newblock \showarticletitle{{Understanding the reproducibility of crowd-reported security vulnerabilities}}. In \bibinfo{booktitle}{\emph{27th USENIX Security Symposium (USENIX Security 18)}}. \bibinfo{pages}{919--936}.
\newblock


\bibitem[OffSec(2024)]%
        {EDB}
\bibfield{author}{\bibinfo{person}{OffSec}.} \bibinfo{year}{2024}\natexlab{}.
\newblock \bibinfo{title}{{Exploit Database}}.
\newblock \bibinfo{howpublished}{\url{https://www.exploit-db.com/}}.
\newblock


\bibitem[Packet(2024)]%
        {PacketSt14:online}
\bibfield{author}{\bibinfo{person}{Packet}.} \bibinfo{year}{2024}\natexlab{}.
\newblock \bibinfo{title}{{Packet Storm}}.
\newblock \bibinfo{howpublished}{\url{https://packetstormsecurity.com/}}.
\newblock


\bibitem[Seebug(2024)]%
        {Seeb31:online}
\bibfield{author}{\bibinfo{person}{Seebug}.} \bibinfo{year}{2024}\natexlab{}.
\newblock \bibinfo{title}{{Seebug}}.
\newblock \bibinfo{howpublished}{\url{https://www.seebug.org/}}.
\newblock


\bibitem[Storm(2024)]%
        {SLMailRemote:online}
\bibfield{author}{\bibinfo{person}{Packet Storm}.} \bibinfo{year}{2024}\natexlab{}.
\newblock \bibinfo{title}{{SLMail 5.1.0.4420 Remote Code Execution}}.
\newblock \bibinfo{howpublished}{\url{https://packetstormsecurity.com/files/161526}}.
\newblock


\bibitem[Suciu et~al\mbox{.}(2022)]%
        {suciu2022expected}
\bibfield{author}{\bibinfo{person}{Octavian Suciu}, \bibinfo{person}{Connor Nelson}, \bibinfo{person}{Zhuoer Lyu}, \bibinfo{person}{Tiffany Bao}, {and} \bibinfo{person}{Tudor Dumitraș}.} \bibinfo{year}{2022}\natexlab{}.
\newblock \showarticletitle{{Expected exploitability: Predicting the development of functional vulnerability exploits}}. In \bibinfo{booktitle}{\emph{31st USENIX Security Symposium (USENIX Security 22)}}. \bibinfo{pages}{377--394}.
\newblock


\bibitem[Sun et~al\mbox{.}(2023)]%
        {sun2023aspect}
\bibfield{author}{\bibinfo{person}{Jiamou Sun}, \bibinfo{person}{Zhenchang Xing}, \bibinfo{person}{Xin Xia}, \bibinfo{person}{Qinghua Lu}, \bibinfo{person}{Xiwei Xu}, {and} \bibinfo{person}{Liming Zhu}.} \bibinfo{year}{2023}\natexlab{}.
\newblock \showarticletitle{{Aspect-level information discrepancies across heterogeneous vulnerability reports: Severity, types and detection methods}}.
\newblock \bibinfo{journal}{\emph{ACM Transactions on Software Engineering and Methodology}} \bibinfo{volume}{33}, \bibinfo{number}{2} (\bibinfo{year}{2023}), \bibinfo{pages}{1--38}.
\newblock


\bibitem[Wang et~al\mbox{.}(2018)]%
        {wang2018revery}
\bibfield{author}{\bibinfo{person}{Yan Wang}, \bibinfo{person}{Chao Zhang}, \bibinfo{person}{Xiaobo Xiang}, \bibinfo{person}{Zixuan Zhao}, \bibinfo{person}{Wenjie Li}, \bibinfo{person}{Xiaorui Gong}, \bibinfo{person}{Bingchang Liu}, \bibinfo{person}{Kaixiang Chen}, {and} \bibinfo{person}{Wei Zou}.} \bibinfo{year}{2018}\natexlab{}.
\newblock \showarticletitle{{Revery: From proof-of-concept to exploitable}}. In \bibinfo{booktitle}{\emph{Proceedings of the 2018 ACM SIGSAC Conference on Computer and Communications Security}}. \bibinfo{pages}{1914--1927}.
\newblock


\bibitem[Ye(2011)]%
        {ye2011cosine}
\bibfield{author}{\bibinfo{person}{Jun Ye}.} \bibinfo{year}{2011}\natexlab{}.
\newblock \showarticletitle{{Cosine similarity measures for intuitionistic fuzzy sets and their applications}}.
\newblock \bibinfo{journal}{\emph{Mathematical and computer modelling}} \bibinfo{volume}{53}, \bibinfo{number}{1-2} (\bibinfo{year}{2011}), \bibinfo{pages}{91--97}.
\newblock


\end{thebibliography}

\end{document}